\documentclass[twocolumn]{aastex63}

\usepackage{graphicx}
\usepackage{epsfig}
\usepackage{times}
\usepackage{natbib}
\usepackage{url}
\usepackage{color}
\usepackage{amssymb,amsmath}
\usepackage{gensymb}
\usepackage{hhline}
\usepackage{comment}

\hypersetup{linkcolor=red,citecolor=cyan,filecolor=cyan,urlcolor=blue}

\newcommand{\p}[1]{{\color{blue}{#1}}}
\newcommand{\q}[1]{{\color{red}{#1}}}

\shorttitle{Solar eruptions triggered by flux emergence}
\shortauthors{T\"or\"ok et al.}

\begin{document}

\title{
Solar Eruptions Triggered by Flux Emergence Below or Near a Coronal Flux Rope
}

\author[0000-0003-3843-3242]{Tibor T\"or\"ok}
\affiliation{Predictive Science Inc., 9990 Mesa Rim Road, Suite 170, San Diego, CA 92121, USA}
\author[0000-0002-4459-7510]{M.~G. Linton}
\affiliation{U.S. Naval Research Lab, 4555 Overlook Ave., SW Washington, DC 20375, USA}
\author[0000-0003-0072-4634]{J.~E. Leake}
\affiliation{NASA Goddard Space Flight Center, Greenbelt, MD 20771, USA}
\author[0000-0002-3164-930X]{Z. Miki\'c}
\affiliation{Retired; performed this work while employed with Predictive Science Inc.}
\author[0000-0001-9231-045X]{R. Lionello}
\affiliation{Predictive Science Inc., 9990 Mesa Rim Road, Suite 170, San Diego, CA 92121, USA}
\author[0000-0001-7053-4081]{V.~S. Titov}
\affiliation{Predictive Science Inc., 9990 Mesa Rim Road, Suite 170, San Diego, CA 92121, USA}
\author[0000-0003-1759-4354]{C. Downs}
\affiliation{Predictive Science Inc., 9990 Mesa Rim Road, Suite 170, San Diego, CA 92121, USA}

\correspondingauthor{Tibor T\"or\"ok}
\email{tibor@predsci.com}

\begin{abstract}
Observations have shown a clear association of filament/prominence eruptions with the emergence of magnetic flux in or near filament channels. Magnetohydrodynamic (MHD) simulations have been employed to systematically study the conditions under which such eruptions occur. These simulations to date have modeled filament channels as two-dimensional (2D) flux ropes or 3D uniformly sheared arcades. Here we present MHD simulations of flux emergence into a more realistic configuration consisting of a bipolar active region containing a line-tied 3D flux rope. We use the coronal flux-rope model of Titov et al. (2014) as the initial condition and drive our simulations by imposing boundary conditions extracted from a flux-emergence simulation by Leake et al. (2013). We identify three mechanisms that determine the evolution of the system: (i) reconnection displacing foot points of field lines overlying the coronal flux rope, (ii) changes of the ambient field due to the intrusion of new flux at the boundary, and (iii) interaction of the (axial) electric currents in the pre-existing and newly emerging flux systems. The relative contributions and effects of these mechanisms depend on the properties of the pre-existing and emerging flux systems. Here we focus on the location and orientation of the emerging flux relative to the coronal flux rope. Varying these parameters, we investigate under which conditions an eruption of the latter is triggered.  
\end{abstract}


%
%
\section{Introduction}
\label{s:intro}
%
The physical mechanisms that initiate and drive coronal mass ejections (CMEs) and associated flares are not yet fully understood. Following \cite{green18}, we consider here as a ``driver'' the mechanism(s) that account for the rapid acceleration and large expansion of the erupting flux observed in CMEs, and as a ``trigger'' any mechanism that can bring a pre-eruptive configuration to a state where the rapid acceleration and expansion of the flux set in. The phase prior to the rapid acceleration, during which some trigger mechanisms may act, is typically associated with the slow rise of coronal loops and/or a filament (if one is present in the CME's source region), as well as with pre-flare signatures, i.e., a gradual increase of EUV and soft X-ray emission \citep[e.g.,][]{chifor07,schrijver08a} and/or a series of small confined flares \citep[e.g.,][]{kliem21}.  

At present, it seems that eruptions are driven by a combination of two mechanisms, namely the ideal magnetohydrodynamic (MHD) torus instability (TI) or ``loss of equilibrium'' \citep[][and references therein]{kliem06,demoulin10,kliem14a} and the (ideal-MHD) consequences of the ``flare reconnection'' occurring below the rising flux rope \citep[e.g.,][]{karpen12}. These two mechanism are closely coupled and support one another \cite[e.g.,][]{vrsnak08a,welsch18}, but it is not yet clear which one dominates under which circumstances (e.g., impulsive eruptions from active regions [ARs] vs.\ eruptions of large quiescent filaments). 

For the trigger, in contrast, it appears that there are a large number of possible mechanisms, both ideal and resistive. Mechanisms that have been suggested are, for example, slow ``tether-cutting'' reconnection \citep{moore92,moore01}, the helical kink instability \citep[e.g.,][]{fan03,torok04a}, magnetic breakout \citep{antiochos99a,macneice04,lynch08}, mass loading \citep[e.g.,][]{low96,seaton11}, magnetic shear reversal \citep{kusano04}, and, more recently, plasmoid formation \citep{roussev12}, solar tornados \citep{su.y12}, flux feeding/transfer \citep{zhang.q14,kliem14}, and the so-called tilt instability \citep{keppens14}; see Table\,1 in \cite{green18} for an extensive compilation of suggested mechanisms.

Many of these trigger mechanisms are a consequence of photospheric flows. For example, flows that converge towards polarity inversion lines (PILs) may induce slow tether-cutting reconnection, which leads to the formation and successive detachment of a flux rope \citep[e.g.,][]{aulanier10}; large-scale shear flows may trigger breakout reconnection higher up in the corona \citep[e.g.,][]{lynch08}; and sunspot rotations may twist up the magnetic field and trigger the kink instability \citep[e.g.,][]{torok03,romano03}.      

The second main process that can eventually lead to CMEs is flux emergence. One may divide the possibilities by which this process produces or triggers an eruption into two scenarios: (i) largely isolated flux that carries its own current, free energy, and self-helicity emerges into quiet-sun fields and produces an eruption after a certain time, when the resulting AR has sufficiently developed, and (ii) the emerging flux merely triggers the eruption of (fully developed) current-carrying coronal flux, i.e., of pre-existing filament channels.

The first scenario has been modeled in a number of idealized numerical simulations spanning the upper convection zone to the low corona, in which the top part of a buoyant sub-photospheric flux rope dynamically emerges into the corona, which in many simulations does not contain a magnetic field. The strong expansion of the emerging flux as it enters the corona leads to the formation of a (line tied) ``envelope'' field overlying the sheared and/or twisted core field. The core field may erupt if it is strong enough to overcome the tension of the envelope field, or if a pre-existing coronal field with an orientation favorable for breakout-like reconnection is present \citep[e.g.,][]{manchester04,archontis08c,archontis10,leake14}. In some simulations recurrent eruptions could be observed \citep{mactaggart09b,archontis14,syntelis17}. On the other hand, if a sufficiently strong pre-existing coronal field with an orientation similar to the orientation of the outer flux surfaces of the emerging rope is present, the rope gets stabilized by the ambient field and evolves to an approximately stable state in the corona \citep[][]{mactaggart11,leake13}. Similar simulations were performed in purely coronal domains, in which the formation of an AR was ``kinematically driven'' by mimicking the rigid emergence of an analytic flux rope at the lower boundary \citep[e.g.,][]{fan03}.    

A variation of this first flux-emergence scenario was presented by \cite{roussev12}. They used a small-scale Cartesian dynamical flux-emergence simulation of the type described above to drive the coronal evolution in a much larger spherical domain, which contained a global dipole field. This was done by rescaling variables extracted from the flux-emergence simulation and imposing them at the bottom boundary of the spherical domain (see Section\,\ref{ss:bc}). The orientation of the emerging flux rope was chosen such that the field lines in its outer flux surfaces were anti-parallel to the dipole field. This led to the formation of a current layer at the interface of the emerging flux and the pre-existing field, and to a ``plasmoid flux rope'' that formed at two null points within the current layer. This constantly growing, highly twisted flux rope finally erupted and evolved into a fast CME. Thus, in this case, it was not merely the emerging flux that evolved and then erupted, but a structure that contained both pre-existing and emerging flux. This constitutes an interesting mechanism for CME formation, but it needs to be explored further \citep[see the discussion in][]{leake14} and to be substantiated with observations. In this respect, we note that \cite{guo.t19} recently demonstrated observational evidence for the formation of a CME from merging plasmoids that were ejected from a vertical current sheet. 

A more complex variation of the first scenario, termed ``collisional shearing,'' was recently suggested by \cite{chintzoglou19}. Here, the collision of opposite-polarity flux concentrations associated with two flux tubes that emerge simultaneously or sequentially within a single AR leads to shearing and flux cancellation and, eventually, to eruptions.  

In the second flux-emergence scenario, eruptions occur when new flux emerges within or close to pre-existing ARs or large quiescent filaments \citep[e.g.,][]{canfield75,williams05,chifor07,sterling07,louis15}. \cite{feynman95} presented a number of cases where the emergence of relatively small-scale (but potentially strong) flux in the vicinity of, or underneath, a filament led to the eruption of the latter in some cases \citep[see also][]{chen00,jing.j04,xu08}. It is reasonable to assume that in such cases eruptions are triggered by some kind of interaction between the emerging and pre-existing flux systems. \cite{feynman95} suggested that, if the orientation of the emerging flux with respect to the large-scale arcade field that harbors the filament is favorable for the formation of a current sheet between the two flux systems, reconnection driven by the ongoing emergence will produce magnetic islands in the sheet, thereby increasing the magnetic pressure in the arcade until it erupts. For flux emerging at the periphery of or close to a filament channel, the orientation was considered as ``favorable'' when the adjacent polarities of the emerging and pre-existing flux were of opposite sign (as in Figure\,\ref{f:fig2}(b),(c)), and as ``unfavorable'' when they were of the same sign (as in Figure\,\ref{f:fig2}(a)). Cases where the new flux emerged within the filament channel were all termed favorable, regardless of their orientation.

\cite{wang99} challenged this view and argued that reconnection driven by emerging flux in the vicinity of a filament channel rather leads to a rearrangement of the field overlying the channel, which then allows the filament to erupt \citep[see also][]{fu16}. They also pointed out that small-scale flux emergence may act as a trigger for eruptions, but is not a necessary condition for their occurrence \citep[see also][]{zhang.y08a}. \cite{ding08} demonstrated by means of 2.5D force-free field calculations that flux emergence near a coronal flux rope leads to the expansion or contraction of the field overlying the rope, depending on the orientation of the emerging flux. Here the flux emergence was represented by a gradual superposition at the photosphere of the normal component of the emerging magnetic flux and the pre-existing normal flux. In their model, a favorable orientation \citep[as defined by][]{feynman95} leads to an expansion of the overlying field, i.e., supporting eruptions, while an unfavorable orientation leads to a contraction of the overlying field, providing additional stabilization to the flux rope and suppressing its eruption. They further showed that such changes of the overlying field occur due to the ideal relaxation of the coronal field to a force-free state following the magnetic connectivity changes imposed by the modified photospheric flux distribution. Reconnection in the coronal field itself was not required to achieve the coronal changes documented in that work, though when it was allowed to occur, it amplified these changes.

\cite{lin.j01} presented an extensive analytical study in which they investigated the response of a 2D coronal flux rope to flux emergence, which was modeled via slowly evolving boundary conditions. They found that the circumstances under which eruptions occur in this seemingly simple system depend in a complicated manner on parameters such as the orientation, strength, and area of the emerging flux, as well as on its distance to the pre-existing flux rope. Specifically, they pointed out that their results suggest that ``there is no simple, universal relation between the orientation of the emerging flux and the likelihood of an eruption,'' but that this needs to be tested in more realistic 3D simulations, which is one of the purposes of our investigation.
 
The first numerical simulations of the second scenario were performed by \cite{chen00} \citep[see also][]{shiota05}. Similar to \cite{lin.j01}, they considered a stable 2D coronal flux rope and kinematically emerged a bipole at the bottom boundary. As in \citet{ding08}, discussed above, they achieved this emergence by gradually superimposing the normal component of the emerging flux and the pre-existing photospheric flux distribution. They investigated four scenarios: emergence next to the rope with an orientation that formed a current sheet between the emerging flux and the pre-existing PIL, emergence next to the rope with an orientation that formed a current sheet on the far side of the emerging flux, and emergence at the PIL with the same or the opposite orientation as the pre-existing photospheric field. 

They found that for emergence next to the rope with the new current sheet forming between the emerging flux and the PIL, the flux rope erupts, while for the inverse orientation it does not, supporting the findings of \citet{feynman95}. For emergence at the PIL, the flux rope erupted when the emerging flux had an orientation opposite to that of the pre-existing photospheric flux, and it was  pulled down toward the photosphere when the emerging flux had the same orientation. This latter result contrasts with the prediction of \citet{feynman95} that emergence at the PIL below a pre-existing filament channel will always cause an eruption. 

\cite{xu05} extended the work by \cite{chen00} to a systematic parameter study and found that the occurrence or lack of an eruption depends not just on the orientation of the emerging flux but also on its amount and on its distance from the PIL, supporting the findings by \cite{lin.j01}. A dependence of eruptive behavior on the location and the amount of emerging flux was also found in 2.5D dynamic simulations of flux emergence into a coronal configuration representing a pseudostreamer \citep{kaneko14}. Simulations of kinematic flux emergence triggering a CME were performed also in large-scale 2.5D spherical (axi-symmetric) domains  \citep[e.g.,][]{dubey06,zuccarello08,zuccarello09a}, where those studies focused mostly on the resulting evolution and properties of the CME. For example, \cite{dubey06} found that the total amount of emerged flux is more important for the final CME speed than the flux-emergence rate.   

Fully 3D MHD simulations of the second scenario, i.e., the triggering of the eruption of a pre-existing, current-carrying coronal structure by localized flux emergence are so far rare. \cite{notoya07} modeled the dynamic emergence of a twisted flux tube from the convection zone into the corona, close to a pre-existing, uniformly sheared magnetic arcade. In their simulation, reconnection between the emerging rope and the arcade leads to the formation of a new flux rope within the arcade which then erupts. The dependencies of  the evolution on parameters such as the shear angle of the arcade or the location, orientation, or amount of the emerging flux were not considered. \cite{kusano12} performed an extensive simulation series of kinematic emergence of a small bipole at the PIL of a uniformly sheared arcade. They varied both the orientation of the bipole and the shear angle (and thereby the free magnetic energy) of the large-scale coronal arcade. Similar to the simulation by \cite{notoya07}, reconnection between the bipole and arcade fields leads in some cases to the formation of a large-scale coronal flux rope within the arcade. \cite{kusano12} found that large-scale, CME-like eruptions occur in their model only for strong arcade-shear ($> 75$\degree) and if the bipole orientation differs by at least 90\degree\, from the normal-polarity orientation of the arcade. They did not, however, consider cases where emergence takes place away from the PIL, e.g., outside of a filament channel.

The purpose of our study described in this article is to extend the existing simulations of the second scenario by considering the effect of flux emergence on a 3D {\em pre-existing coronal flux rope}. Our simulations are driven kinematically, and we impose the emerging flux at the PIL, as well as at different distances from the PIL, and we vary its orientation. The variations in distance and orientation of the emerging flux are chosen to emulate the circumstances reported on by \citet{feynman95}, namely, flux emergence at or near the PIL with orientations that are either favorable or unfavorable for reconnection, in the terminology of \citet{feynman95} (see Table \ref{t:tab1} and associated discussion below). In contrast to the ad-hoc descriptions of flux emergence used in previous simulations of the second scenario \citep[except for][]{notoya07}, we employ a more self-consistent approach by taking our flux-emergence electric fields directly from a dynamical flux-emergence simulation.

%
%
\section{Numerical Setup}
\label{s:num}

For the simulations described in this article, we employ the MHD code ``Magnetohydrodynamic Algorithm outside a Sphere'' (MAS; Section\,\ref{ss:MAS}) to model the emergence of a compact, strong bipolar AR into a corona that contains a much larger but weaker pre-existing coronal flux rope (see Figure\,\ref{f:fig1}(a)). The evolution is driven at the lower boundary of the coronal domain using data that were extracted from a dynamic flux-emergence simulation performed with the MHD code Lare3D (Section\,\ref{ss:lare}). In this section we describe the our simulation setup, paying particular attention to the coupling of Lare3D and MAS. We note that a similar coupling (in thermodynamic MHD, whereas the runs presented here are in the $\beta=0$ approximation) was recently used to model the formation and propagation of coronal jets \citep{torok16} and to estimate their contribution to the solar wind \citep{lionello16}. Furthermore, Runs 6 and 7 (see Table\,\ref{t:tab1}), and a simulation similar to Run 5 were used in \cite{dacie18} to interpret the splitting and subsequent eruptions of a ring-shaped, quiescent filament.

\subsection{Coronal Simulations (MAS)}
\label{ss:MAS}
%
The simulations presented here were performed using the MHD code MAS, developed and maintained at Predictive Science Inc. \citep[e.g.,][]{mikic94,mikic99,lionello99,linker01,lionello09,caplan17,torok18}, which advances the standard viscous and resistive MHD equations. The $\beta=0$ approximation, in which thermal pressure and gravity are neglected, is employed, so that the evolution is driven entirely by magnetic and inertial forces. The use of this approximation is justified here, since the dynamics relevant for our investigation occur in the low corona, where the plasma $\beta$ (the ratio of thermal to magnetic pressure) is low \citep{gary01}. 

For the $\beta=0$ approximation, the MHD equations reduce to the following form of the momentum and the induction equations, respectively \citep[see][]{lionello02}:

\begin{equation}
    \rho(\partial\mathbf{v}/\partial t + \mathbf{v}\cdot\nabla\mathbf{v})
    = \mathbf{J}\times\mathbf{B}/c + \nabla\cdot (\rho\nu\nabla\mathbf{v}),
    \label{eq:momentum}
\end{equation}

\begin{equation}
    \partial\mathbf{A}/\partial t = -c\mathbf{E} = \mathbf{v}\times\mathbf{B}-c\eta\mathbf{J}.
    \label{eq:induction}
\end{equation}

Note that MAS evolves the vector potential $\mathbf{A}$. The magnetic field $\mathbf{B}$ and the electric current $\mathbf{J}$ are then expressed as:
\begin{equation}
    \mathbf{B}=\nabla\times\mathbf{A};
    \mathbf{J}=c\nabla\times\nabla\times\mathbf{A}/4\pi.
\end{equation}
The induction equation makes use of the electric field $\mathbf{E}$, which has the form:
\begin{equation}
    \mathbf{E} = -\mathbf{v}\times\mathbf{B}/c + \eta\mathbf{J}.
\end{equation}

The MHD equations are calculated in dimensionless form, where lengths, times, field strengths, and mass densities are normalized by $R_\odot$=696\,Mm, $\tau_A$=1445.87\,s ($\approx 24$\,min), 2.2069\,G, and 1.6726$\times 10^{-16}$\,g\,cm$^{-3}$, respectively, where $R_\odot$ is the solar radius and $\tau_A$ is the Alfv\'en time. A uniform resistivity is chosen such that the Lundquist number $\tau_R/\tau_A$ is $5.0 \times 10^6$, where $\tau_R$ is the resistive diffusion time. A uniform viscosity $\nu$ is used, corresponding to a viscous diffusion time $\tau_{\nu}$ such that $\tau_{\nu}/\tau_A = 500$. This value is chosen to dissipate unresolved scales without substantially affecting the global solution. In the coronal flux rope, where the characteristic length and the Alfv\'en speed substantially differ from the above values (see \S\,\ref{ss:TDm}), the Lundquist number has a similar value of $2.7 \times 10^6$, with the same ratio of resistive and viscous diffusion times. We note that when dynamics such as reconnection begin in the simulation, the numerical diffusion dominates the specified resistivity.

We use a nonuniform spherical ($r,\theta,\phi$) mesh of size $258\times 461\times 561$ that ranges from $r=1.0$ (the solar surface) to $3.5$. We use a radial resolution of $\Delta r=0.0005$ (0.35 Mm) in the height range $r=1.0-1.075$ (extending above the apex of the flux-rope axis; see Section\,\ref{ss:TDm}), increasing to $\Delta r=0.11$ at the outer boundary. We use a latitudinal/longitudinal resolution of 0.0005 radians in an area covering 0.125 and 0.14 radians in $\theta$ and $\phi$, respectively, within which the coronal flux rope is located and the flux emergence is imposed. The resolution of the latitudinal/longitudinal cells outside of this area then decreases to 0.05 radians. We note that, even though the lower boundary of the MAS domain is associated with the solar surface, it should physically be considered here as the bottom of the corona, since we use the $\beta=0$ approximation.

Closed-wall boundary conditions are used at the outer radial boundary of the simulation, which is justified since we are interested here only in the initiation of CMEs low in the corona. The inner radial boundary, representing the base of the solar corona, is driven by setting the velocity and electric fields at that boundary, $\mathbf{v_0}$ and $\mathbf{E_0}$, to input velocities and electric fields which are chosen so as to produce the desired boundary evolution. Section \ref{ss:bc} describes the method used to set this boundary condition from the Lare3D data.

\subsection{MAS Initial Configuration (TDm)}
\label{ss:TDm}

The initial coronal magnetic field consists of a flux rope embedded in a bipolar AR (see Figures\,\ref{f:fig3}(a) and \ref{f:fig4}(a) below). The AR is constructed by means of a fictitious horizontal dipole \citep[see, e.g.,][]{titov08}, which is placed at $r=0.96$ (in units of $R_\odot$ [see \S\,\ref{ss:MAS}], i.e., 28 Mm below the solar surface), with the dipole moment vector pointing towards the north pole and the dipole strength chosen such that the maximum $|B_r|$ at the surface is 89\,G. We then employ the modified Titov-D\'emoulin model \citep[TDm;][]{titov14} to embed a flux rope into the AR, such that the field of the AR, often called the `strapping field,' restrains the flux-rope field, and so the flux rope is in a stable magnetic equilibrium. To this end, the geometrical axis of the (toroidal) TDm current channel is aligned with the PIL of the AR (i.e., the $\phi$ coordinate), with its apex located exactly above the AR center. For the TDm torus we use a minor radius of $a=0.045$ (31 Mm) and place it such that the distance between the two foot points of the torus axis is 114 Mm, and its apex is 35 Mm above the surface. The total length of the resulting flux rope (as projected on the surface) is $\approx$\,175\,Mm. The TDm axial current, $I$, is set to its equilibrium value, $I_S$ \citep[see][]{titov14}. In the $\beta=0$ approximation the distribution of the mass density can be chosen freely. Here we use a fixed radial profile $\rho(r)$ that was derived from a thermodynamic MHD solution of the corona \citep[e.g.,][]{lionello09}.  

We then relax this configuration for one Alfv\'en time to obtain a sufficiently well relaxed numerical equilibrium. During the relaxation, the initially circular current channel adopts a ``teardrop'' shape (see Figure\,\ref{f:fig3}(a)), as it adjusts to the bipolar background field \citep[see][]{titov14}. After the relaxation, the apex of the magnetic axis of the TDm flux rope is located at $r=1.062$, corresponding to $\approx 43$\,Mm above the lower boundary (cf. the top panel in Figure\,\ref{f:fig4}(a)). The relaxed rope has a diameter of $\approx 60$\,Mm (in the radial direction) and the Alfv\'en speed at the apex of the rope's axis is $\approx 3,000\,\mathrm{km\,s}^{-1}$, yielding an Alfv\'en crossing time of $\approx 20$\,s. The flux emergence is then imposed starting at $t=1$ (in units of $\tau_A$; see \S\,\ref{ss:MAS}). For all simulations described here, we impose the emergence until $t=14.5$ (unless stated otherwise).

%
\begin{figure}[t]
\centering
\vspace{1.5mm}
\includegraphics[width=1.\linewidth]{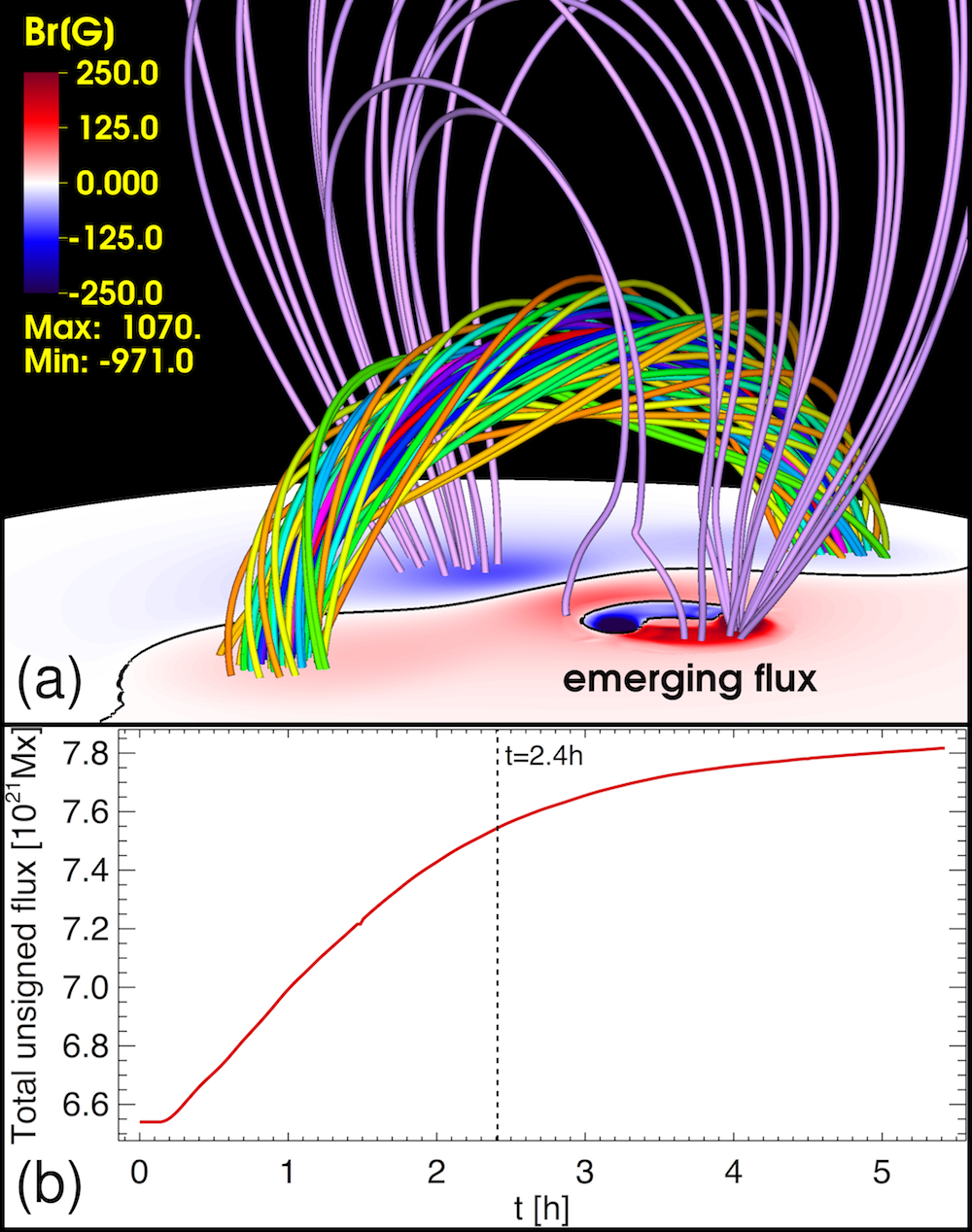}
\caption{
(a) Example configuration of the simulations described here. The bottom plane shows the radial magnetic field at the solar surface, $B_r(r=1)$. A compact bipolar flux region emerges next to a large pre-existing coronal flux rope (colored field lines). Purple field lines show the ambient field surrounding the coronal flux rope, some of which has reconnected with the emerging flux. This configuration is from Run 6 (see Table\,\ref{t:tab1}), shown at $t=7$ (2.4 hours after the onset of flux emergence). 
(b): Magnetic flux evolution at $r=1$. The vertical line marks the time shown in (a). 
}
\label{f:fig1}
\end{figure}

\vspace{8mm}

\subsection{Lare3D Flux Emergence Simulation}
\label{ss:lare}
%
After the relaxation phase, we model the emergence of a compact bipolar AR into the pre-existing coronal TDm configuration to initiate interactions with the existing configuration. To make the emerged fields as realistic as possible, we used a simulation of the emergence of a magnetic flux rope into a magnetic-field--free corona from beneath the photosphere \citep{leake13} as a source for the driving boundary conditions in MAS (see next section). That simulation was performed in Cartesian coordinates using the Lare3D MHD code \citep{arber01}, producing a  coronal configuration that mimics a small isolated bipolar AR with a flux rope that is stabilized by its own envelope field. The simulation used here is similar to the cases ``ND'' and ``ND1'' described in \cite{leake13}.

In all simulations described in this paper, we fix the center of the emerging AR at the same latitude as the center of the TDm configuration, and vary only its longitudinal ($\theta$) position, i.e., the distance of the emerging flux rope to the TDm configuration. Figure\,\ref{f:fig1}(a) shows an example, in which the newly emerging magnetic field is labeled as ``emerging flux.'' This simulation is described in detail in Section\,\ref{ss:counter-j_co-B}. In Figure\,\ref{f:fig1}(b) we plot the evolution of the emerging flux in this simulation at $r=1$. The AR emerges for about 1.5 hours at an almost constant rate of about $5 \times 10^{20}$\,Mx per hour, after which the emergence gradually slows down. After six hours, when the emergence has essentially saturated, the total unsigned AR flux is $\approx$\,1.3 $\times 10^{21}$\,Mx, which corresponds to about 20 per cent of the total flux of the TDm configuration. At this time, the emerging bipole has reached a size of $\approx$\,50\,Mm.

\subsection{MAS-Lare3D Coupling (Boundary Conditions)}
\label{ss:bc}
%
Our goal here is to emerge a small magnetic AR through the bottom boundary of MAS into the simulated MAS atmosphere. While the simplest way to do this is to kinematically transport a pre-defined magnetic shape up through the bottom boundary, as was done, for example, in the flux emergence simulations by \citet{fan03} and \citet{kusano12}, it is known that this does not produce a realistic flux emergence evolution. As has been shown in fully dynamic flux emergence simulations \citep[e.g.,][]{fan01,archontis04,manchester04}, some of which also included realistic convection \citep[e.g.,][]{fang10}, the dramatic stratification at and near the photosphere has a strong distorting effect on emerging magnetic fields, and in fact this stratification usually splits up and completely reconfigures emerging magnetic flux - in some sense the exact opposite of simple vertical translation. Therefore, we choose the more involved method of taking our driving conditions for the bottom boundary of MAS from a time series of slices extracted from the fully dynamical Lare3D simulation of flux emergence, which self-consistently produces this splitting and rearrangement of the emerging flux. 
 
The boundary electric field used to drive the simulation (see below) is calculated from the Lare3D slices, assuming that resistivity in the slice is negligible, so that the $\eta\mathbf{J}$ term in $\mathbf{E}$ can be dropped. Here an approximation must be made to account for the fact that the Lare3D simulation emerges flux into an initially field-free atmosphere, while the initial MAS atmosphere contains the field of the pre-existing TDm configuration. We approximate this combined boundary by superimposing the evolution of the Lare3D field onto the pre-existing MAS field as described below. While this will not give exactly the same evolution as would be expected from a fully dynamical simulation of such a configuration, the resulting simulations show that in practice, a realistic twisted AR emerges dynamically into MAS, thus creating the physical effect needed to carry out this set of investigations. 
 
A number of investigations reported in the literature have driven boundary electric fields to energize pre-existing coronal field, but with the exception of \cite{roussev12}, all have used ad-hoc forms for the driving electric field. For example, previous work with MAS has investigated the introduction of sheared field, i.e., magnetic field aligned with the PIL, into the corona as a driver of CME eruptions \citep[e.g.,][]{mikic13}. In those investigations, the electric field which introduced this ad-hoc shear was constrained so that it preserved the photospheric contours of $\mathbf{B}$ normal to the boundary. This property of invariance was desirable primarily because it preserved the input radial magnetogram, which for many cases studied by MAS was taken from a specific observation being modeled. It was also desirable because there was no readily available analytical description of what the electric field actually should look like for true flux emergence, and this method allowed for a simple description of the electric potentials in terms of the pre-existing normal field $B_r$ and the desired amount of shear to be introduced. \citet{zuccarello18} used a similar method to energize a coronal arcade with the introduction of shear. Note that in both of these simulations, eruptions were not generated directly from the introduction of shear, but from subsequent concentration and cancellation of that shear by imposed flows converging towards the PIL. Similarly, a number of simulations run with the ARMS code \citep[e.g.,][]{antiochos99a, lynch08} impose shear fields close to the PIL via a prescribed photospheric flow field which preserves $B_r$. This effectively acts as an imposed electric field and, again, this gives an easily formulated recipe for generating shear fields in a simulation. 

The flux emergence boundary driving method we will use here addresses many of the issues with these earlier ad-hoc methods. In particular, it explicitly includes the increase in normal flux penetrating the photosphere and the normal velocity which transports this flux through the photosphere. Both of these are classic components of flux emergence, but are purposely left out of the simulations discussed above which focus on the introduction of shear to energize the corona \citep[e.g.,][]{antiochos99a,mikic13, zuccarello18}. In addition, it explicitly includes the horizontal expansion flows and the topological rearrangement of the emerging field generated by dynamical flux emergence which are left out of the flux emergence conditions such as those imposed in \cite{chen00,fan03,kusano12}. For the simulations we carry out here, which are focused explicitly on what the emergence of magnetic flux will do to a pre-existing bipolar AR that contains a flux rope, it is critical that these two dominant features of flux emergence be explicitly included in as self-consistent a manner as is feasible.

The Lare3D simulations were carried out in Cartesian $(x,y,z)$ coordinates in a small region of the solar surface, with $z$ being the vertical direction.  To develop the boundary driving electric field for MAS, we first extract the driving velocity, $\mathbf{v_{Lare3D}^0}$, and magnetic field, $\mathbf{B_{Lare3D}^0}$, from the plane at $z=24$ (0.68 Mm above the nominal photosphere at $z=20$)\footnote{In the Lare3D simulation used here the ``photosphere/chromosphere'' layer is located at $z=[20,30]$, while in \cite{leake13} it was located at $z=[0,10]$. Apart from this shift along the $z$ coordinate, the initial atmospheric stratification is the same as in \cite{leake13}.} in Lare3D as a sequence in time. We project these onto a small section of the spherical MAS surface at $r=R_0$ by mapping the vectors in this plane onto the spherical surface. Note that this is different from \cite{roussev12}, who imposed the extracted data on a flat plane embedded in their 3D spherical boundary. 

We choose to evolve the boundary in a way where the dynamic emergence of radial magnetic flux simulated in Lare3D is superimposed on the initial TDm radial flux at the boundary. In a truly dynamic flux emergence simulation, the form of the emerging flux would not only distort as it does in the Lare3D model, but it would also push aside and distort the TDm flux. However, this would require a full dynamical simulation of the convection zone. We choose a compromise where the emerging flux is itself distorted by the emergence (as simulated by Lare3D) but is then simply superimposed on the radial TDm flux. The transverse TDm flux, in contrast, is allowed to evolve due to the imposed Lare3D electric currents. 

The boundary velocity and magnetic field for the electric field driver are therefore set to:
\begin{equation}\label{eqn:velocity}
    \mathbf{v^0}=\mathbf{v^0_\text{Lare3D}}/f,
\end{equation}
\begin{equation}
         \mathbf{B^0} = \mathbf{B}^0_\text{Lare3D}+\mathbf{\hat{r}}B^0_\text{r,TDm},
\end{equation}
where $f=2.5$ is a scale factor used to slow the emergence timescale down to a more realistic time scale. As pointed out in the literature \citep[e.g.,][]{norton2017}, simulated flux emergence is often significantly faster than observed flux emergence. The factor of $f=2.5$ used here is chosen to generate a smoother emergence evolution and more realistic emergence speed.

We then calculate the driven time evolution of the vector potential at the boundary using the driving electric field as follows. The normal ideal electric field from Lare3D and the normal electric current at the boundary from the evolving MAS state, $J^0_r$, are used to advance the normal component of the vector potential:
\begin{equation}
\frac{1}{c}\frac{\partial A^0_r}{\partial t}=-E^0_r
=\mathbf{\hat{r}}\cdot(\mathbf{v^0}\times\mathbf{B^0})-\eta J^0_r.
\end{equation}
The tangential components of the vector potential are advanced by the equation
\begin{equation}
\frac{1}{c}\frac{\partial\mathbf{A^0_t}}{\partial t}=-\mathbf{E^0_t}=-\mathbf{\nabla_t}\times\Psi\mathbf{\hat{r}}-\mathbf{\nabla_t}\Phi,
\label{eq:At-bc}
\end{equation}
where transverse electric field $\mathbf{E}^0_t$ is decomposed into the two potentials $\Psi(\theta,\phi)$ and $\Phi(\theta,\phi)$, as in \cite{lionello13a}. This treatment of the transverse electric field represents a form of ``divergence cleaning'' that attempts to preserve the transverse 
curl and transverse gradient components of the electric field in the $r=R_0$ surface). 

The potential $\Psi$ is found from the evolution of the radial magnetic field at the boundary:
\begin{equation}
\frac{1}{c}\frac{\partial B^0_r}{\partial t}=\nabla^2_t\Psi.
\end{equation}
This elliptic equation is solved for $\Psi$ in the surface $r=R_0$ for the evolution of $B^0_r$ taken from the Lare3D simulation. Note that the radial field of the intitial TDm configuration is included in this calculation - but it does not contribute to $\Psi$, as it has no time derivative. The $\Phi$ potential is advanced by taking the divergence of Equation (\ref{eq:At-bc}), to give
\begin{equation}
-\nabla^2_t\Phi=-\mathbf{\nabla_t}\cdot\mathbf{E^0_t}=\mathbf{\nabla_t}\cdot\left(\mathbf{v^0}\times\mathbf{B^0}\right)_t.
\label{eq:Phi-eq}
\end{equation}
This elliptic equation is solved for $\Phi$ in the surface $r=R_0$. The combination of $\Psi$ and $\Phi$ then define the transverse electric field $\mathbf{E^0_t}$ and used to advance $\mathbf{A^0_t}$ according to Equation\,(\ref{eq:At-bc}). This scheme has the benefit that the evolution of $B^0_r$ is reproduced exactly. 

It is to be noted that at the bottom boundary of the simulations reported in this article, the maximum strength of the fully-emerged radial bipole field is one order of magnitude larger than the maximum strength of the pre-existing TDm radial field, since we want the emerged bipole to perturb the system significantly.

The bottom boundary conditions for the momentum equation in MAS (Equation \ref{eq:momentum}) 
are simply set so that the velocity at that boundary is the same as that used to calculate the driving electric fields, i.e., these are set by Equation (\ref{eqn:velocity}).

\begin{figure}[t]
\centering
\includegraphics[width=1.\linewidth]{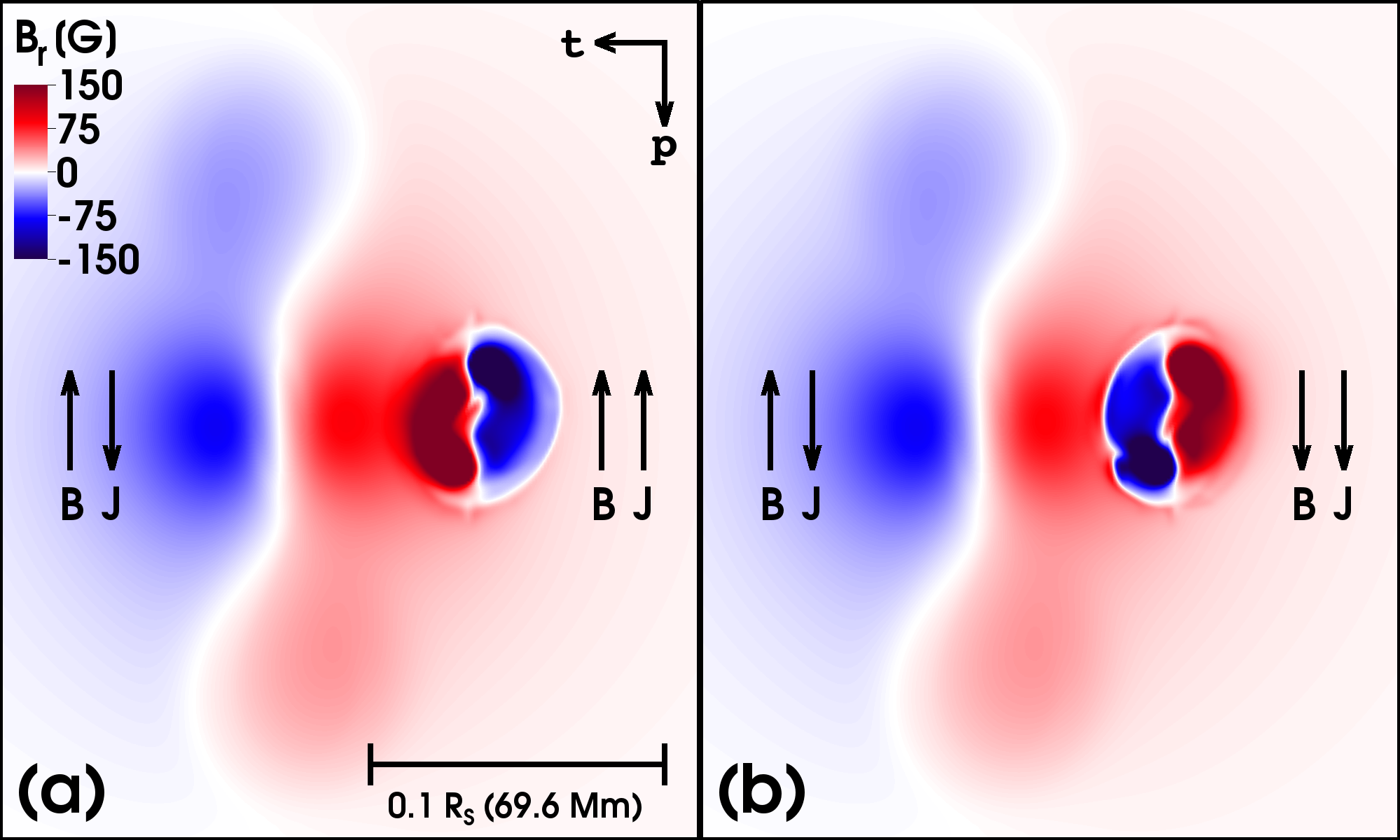}
\caption{
Photospheric magnetic field configuration for two representative simulations (cf. Figure\,\ref{f:fig1}(a)). Both panels show $B_r(r=1)$ at $t=7$ (2.4 hours after the onset of flux emergence). Panel (a) shows Run 2 (see Section\,\ref{ss:counter-j_co-B}), for which the configuration is unfavorable for reconnection in the paradigm of \citet{feynman95}. Panel (b) shows Run 6 (see Section\,\ref{ss:co-j_counter-B}), for which the configuration is favorable for reconnection in that paradigm. For context, within each panel, arrows indicate the respective orientations of the axial component of the current (J) and the axial component of the magnetic field (B) for the TDm flux rope (left) and the emerging flux rope (right). 
The distance between the AR centers is 45\,Mm in these cases. The $\theta$ and $\phi$ axis orientations (t and p, respectively) and the length scale ($R_S$ is the solar radius) are shown in panel (a).
}
\label{f:fig2}
\end{figure}

This philosophy of driving flux emergence through the bottom boundary of a coronal simulation by using electric fields from a separate, dynamical flux emergence simulation was introduced by \cite{roussev12}. Similarly to what we have just described, they drove the emergence by extracting the velocity and magnetic field from such a simulation \citep{archontis04} and using this to calculate a boundary-driving electric field. However, as their goal was to mimic a large-scale AR emergence, but they only had a small-scale emergence simulation to work with, they multiplied the spatial scales of the emerging region by a factor of $L=10$. In addition, as they extracted their fields from the photosphere of the flux emergence simulation, but introduced them at the low $\beta$ boundary of the coronal simulation, they multiplied their magnetic pressure $B^2$ by a factor of $D=1/5,600$ over a four-grid cell boundary layer below the coronal boundary. It is not clear how this may have changed the realism of the flux-emergence condition, but a simple analysis shows that this would have increased the total flux by a modest factor of $L^2\sqrt{D}=1.3$, and would have decreased the local Lorentz forces by the significant factor of $D/L=1/56,000$. To preserve the self-consistency of our boundary conditions as much as possible, we preserve both the input length scales and the input field strength in our simulations. In the flux-emergence simulation we are using to drive our simulations, the emerging flux significantly expands once it enters the less dense atmosphere. This means that the horizontal extent of the emerging flux increases with height above the photosphere, while the Lorentz forces are decreasing with height. Therefore, to mitigate the effect of unbalanced Lorentz forces in our simulations, while at the same time ensuring sufficient spatial resolution to resolve the emerging flux, we extract the magnetic fields and velocities from a plane four photospheric pressure scale heights above the photosphere.

270 slices are used, spaced at 72s. The velocity and magnetic fields in the input slices was smoothed spatially using a low-pass filter and linearly interpolated between input times. For each slice of the sequence, the input Lare3D data is masked out beyond the central emergence area: velocity and magnetic fields in pixels with $B < 6\times 10^{-2} B_\mathrm{max}$, where $B_\mathrm{max}$ is the maximum total field strength within the slice, are set to zero, to remove low-amplitude fine-scale magnetic structures at the periphery of the emerging flux area. 
  
%
\begin{figure*}[t]
\centering
\includegraphics[width=1.\linewidth]{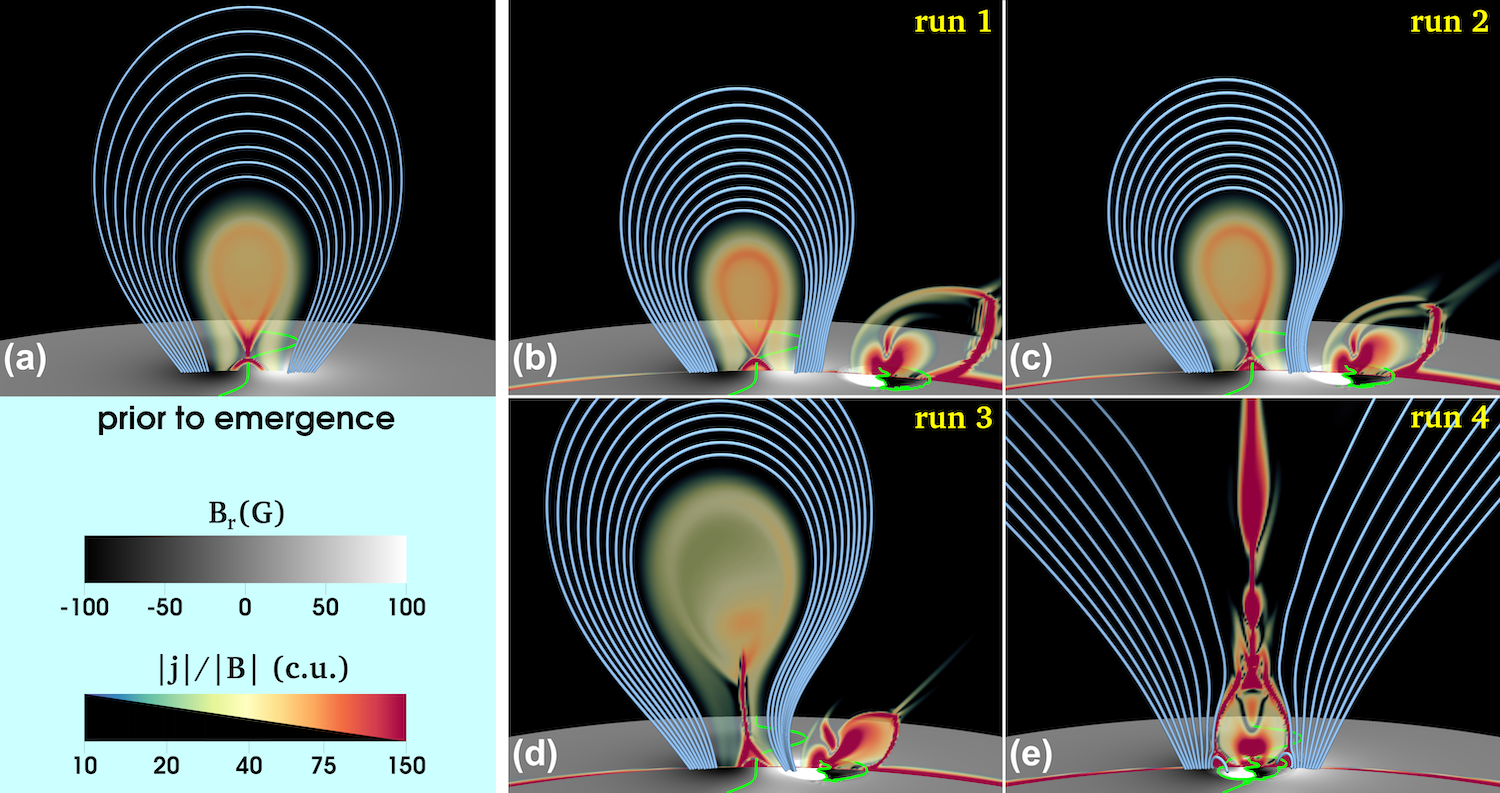}
\caption{
Summary of Runs 1--4. The evolution is visualized by $|{\bf j}|/|{\bf B}|$ (in code units) in a transparent layer cutting through the flux-rope centers, and by field lines overlying the TDm rope. The field lines start at the same positions on the left-hand side of the TDm rope in all panels. The green lines outline PILs.
(a) TDm configuration after relaxation ($t=1$). 
(b--e) Runs 1--4, all shown at $t=10$ (about 3.6 hrs after the onset of flux emergence). In Run 4 the erupting TDm rope has already left the field of view at this time. 
}
\label{f:fig3}
\end{figure*}

%
%
\section{Results}
\label{s:res}

\begin{table}[t]
\centering
\begin{tabular}{|c| c | l | r | r | l|}
\hline\
No.    & 
FM95 &
$j_\mathrm{axial}$ & 
$B_\mathrm{axial}$ &
$\Delta_\mathrm{FR}$ &   
TDm Evolution \\
%
\hhline{|=|=|=|=|=|=|}
1      & U & counter & co      & 59 &   No eruption \\ 
\hline
2      & U & counter & co      & 45 &   No eruption \\ 
\hline
3      & U & counter & co      & 31 &   Eruption    \\ 
\hline
4      & F & counter & co      &  0 &   Eruption    \\ 
%
\hhline{|=|=|=|=|=|=|}
%
5      & F & co      & counter & 59 &  Eruption (D) \\ 
\hline       
6      & F & co      & counter & 45 &  No eruption (S) \\ 
\hline
7      & F & co      & counter & 31 &  Splitting \\ 
\hline
8      & F & co      & counter &  0 &  Splitting \\ 
\hline
\end{tabular}
\caption{
Summary of runs. The first column shows the run number, the second column denotes whether the configuration is ``favorable'' (F) or ``unfavorable'' (U) for reconnection according to the classification in \cite{feynman95}, $j_\mathrm{axial}$ ($B_\mathrm{axial}$) denotes whether the axial currents (axial magnetic fields) of the pre-existing TDm flux rope and the emerging Lare3D flux rope are co-aligned or counter-aligned, and $\Delta_\mathrm{FR}$ is the distance between the centers of the TDm and Lare3D flux systems on the solar surface (in Mm). The last column summarizes the resulting evolution of the TDm rope. The letters in parentheses indicate that the eruption of the TDm rope was ``delayed'' (D) or ``suppressed'' (S) by the erupting Lare3D rope (see Section\,\ref{ss:co-j_counter-B} for details).     
}
\label{t:tab1}
\end{table}

In this section we describe the results of our simulations. We investigate eight scenarios, summarized in Table \ref{t:tab1}, varying both the location of the emerging magnetic field and its orientation relative to the TDm field. For Runs 1--4 the Lare3D flux rope, prior to emergence, is oriented such that the axial component of its magnetic field is parallel to the axial component of the TDm rope's magnetic field (``co-aligned'') and the axial component of its current is anti-parallel to the axial component of the TDm rope's current (``counter-aligned''). In contrast, Runs 5--8 have the inverse orientation for the emerging flux, giving counter-aligned axial magnetic fields and co-aligned axial currents. 

Four distances between the center of the TDm configuration at the surface and the center of the emerging flux are examined: 59 Mm (Runs 1 and 5), 45 Mm (Runs 2 and 6), 31 Mm (Runs 3 and 7) and 0 Mm (Runs 4 and 8). As illustrated in Figure \ref{f:fig2}, this means that Runs 1--3 (Runs 5--7) are unfavorable (favorable) for reconnection in the paradigm of \citet{feynman95}. Both cases in which emergence occurs directly under the TDm rope, i.e., Runs 4 and 8, are denoted favorable in that paradigm.

For cases where the new flux emerges relatively far from the TDm rope, we expect that it will interact predominantly with the bipolar AR strapping field, and we study to which extent such interactions may affect the stability of the TDm rope. In these cases, the orientation of the emerging field relative to the orientation of the strapping field 
should be the primary characteristic controlling the evolution, as it  determines on which side of the emerging flux region a current layer forms and reconnection takes place, i.e., at which height above the TDm rope the strapping field will be affected.  

For cases where the Lare3D flux rope emerges sufficiently close to the TDm flux rope, we expect a direct interaction of the ropes, in the form of attraction/repulsion of the flux-rope currents and reconnection between their magnetic fields. For such interactions, both the relative directions of the axial currents and of the axial magnetic fields should play an important role. \cite{linton01} performed numerical experiments in which they forced two straight, {\em sub-photospheric} flux ropes to collide. They found that ropes with counter-aligned axial currents repelled each other, an interaction they termed a ``bounce.'' No further interaction occurred following a bounce, so for those cases, the direction of the axial magnetic field was unimportant. On the other hand, flux ropes with co-aligned axial currents attracted each other. This led to strong reconnection between them, and so their axial-field directions became important as well. Two distinct behaviors were found in these cases. For ropes with co-aligned axial magnetic fields, those fields simply reinforced, and a new, single flux rope was formed. This type of interaction was termed a ``merge'' by \cite{linton01}. In contrast, for ropes with counter-aligned axial magnetic fields, those fields reconnected, and the two colliding ropes split into two new flux ropes in an interaction termed a ``slingshot.'' Slingshot reconnection between two {\em coronal} flux ropes that were driven towards each other by photospheric flows was found by \cite{torok11} in a simulation that was designed to reproduce an observed footprint-connectivity exchange of two adjacent filaments \citep{chandra11a}. 

We note that in our setup the flux ropes have opposite signs of twist: the TDm rope is left-handed and the Lare3D rope is right-handed. This means that either the axial currents are co-aligned and the axial magnetic fields are counter-aligned, or vice versa. Thus, for the studies to be presented here, we expect to see the equivalent of axial-current repulsion (bounce) and axial-current attraction plus axial-field reconnection (slingshot) interactions. If we were to study cases where both flux ropes had the same sign of twist, we could similarly expect either axial-current repulsion (bounce), or axial-current attraction followed by axial-field reinforcement (merge). For the cases in our study where rope-rope interactions happen, the primary interaction dynamics of interest are expected to be the axial current repulsion or attraction, rather than whether attraction leads to merge or slingshot reconnection. Thus, to keep the size of our investigation manageable, we limit it to opposite-twist configurations and leave the cases of like-twist configurations to future investigations.

\subsection{Counter-aligned axial current \& co-aligned axial field}
\label{ss:counter-j_co-B}
\begin{figure*}[t]
\centering
\includegraphics[width=1.\linewidth]{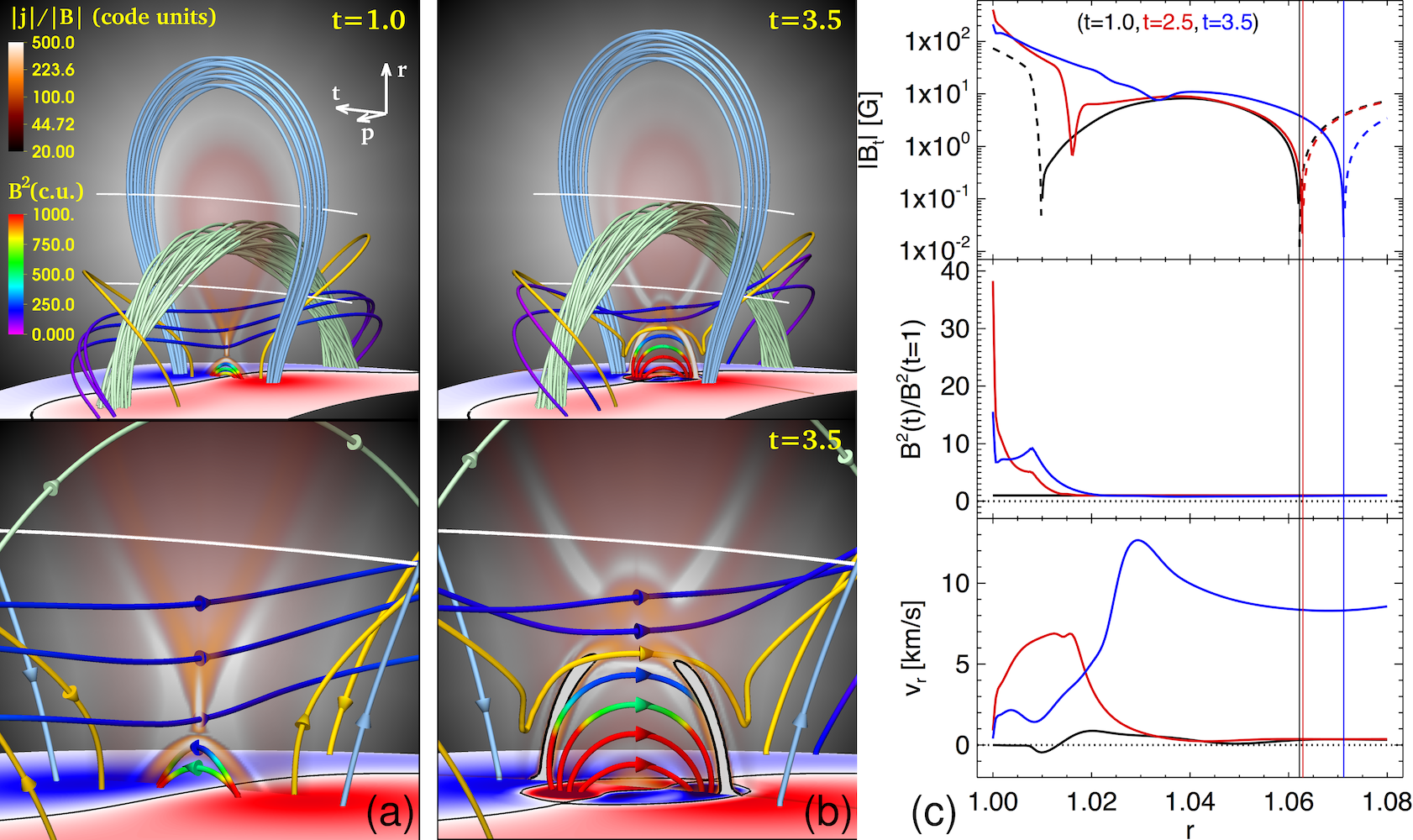}
\caption{
Early evolution in Run 4. 
(a) System at $t=1$, prior to flux emergence (cf. Figure\,\ref{f:fig3}(a)). Light-green (light-blue) field lines show the core of the TDm rope (the overlying field). Field lines below the rope core are colored by the magnetic energy density, $B^2$ (in code units), except for the reconnecting orange field lines (see text for details). The magnetogram is as in Figure\,\ref{f:fig2}; black lines are PILs. Electric currents are visualized by $|{\bf j}|/|{\bf B}|$ (in code units) in a transparent layer across the rope center. White lines indicate heights $r=1.04$ and 1.08, respectively. The bottom panel magnifies the area below the TDm rope core; arrows indicate field directions. For clarity, both the rope core and the overlying field are depicted by only one field line.
(b) Same as (a) at $t=3.5$. The foot points of the orange field line are the same as in (a); the starting points of all other field lines are chosen in the volume to best represent the respective flux systems. Current layers at the edge of the emerging flux are highlighted by black contours of 0.08\,$(|{\bf j}|/|{\bf B}|)_\mathrm{max}$ in the bottom panel. 
(c) Top to bottom: Magnetic field component perpendicular to the TDm rope axis (dashed lines are negative values); magnetic energy density relative to its value at $t=$1; radial flow velocity, all shown as a function of height at the $(\theta,\phi)$ location of the center of the TDm rope and the emerging flux, at times $t=$1.0 (black), 2.5 (red), and 3.5 (blue). Vertical lines mark the respective positions of the rope axis.
}
\label{f:fig4}
\end{figure*}

\begin{figure*}[t]
\centering
\includegraphics[width=.93\linewidth]{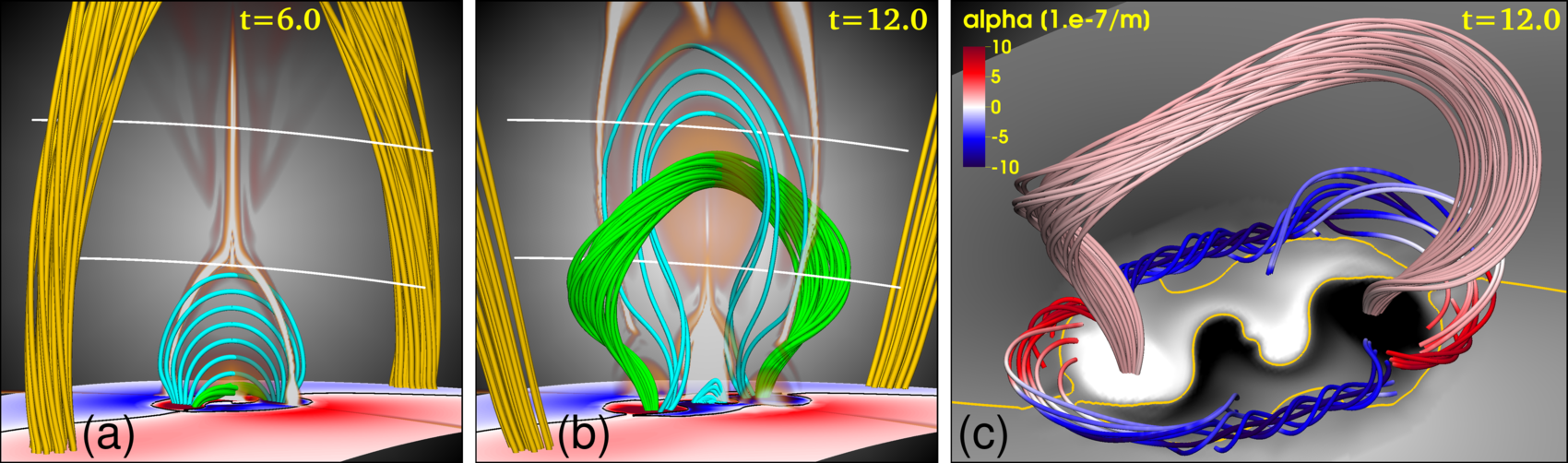}
\caption{
Eruption of the Lare3D flux rope in Run 4.
(a) Configuration at $t=6$ (orange field lines show the erupting TDm rope; cf. top panels in Figure\,\ref{f:fig4}(a) and (b)). The emerging flux has expanded (cyan field lines); a small twisted core is visible just above the surface (green field lines).
(b) Eruption of the emerging rope at $t=12$. Green field lines show the core of the rope; cyan field lines show outer and reconnected flux.
(c) Oblique view of the configuration at the same time as in (b). Field lines are colored by the force-free parameter $\alpha={\bf j} \cdot {\bf B}/B^2$. $B_r(r=1)$ is shown in gray scale. Two pairs of ``conjoined flux ropes'' have formed above the PIL surrounding the emerging flux (see text for details).
}
\label{f:fig5}
\end{figure*}

%
We first consider configurations in which the axial currents of the pre-existing TDm flux rope and the emerging Lare3D rope are counter-aligned (oppositely directed), while their axial magnetic fields are co-aligned (pointing in the same direction), as visualized in Figure\,\ref{f:fig2}(a). These runs are labeled 1--4 (see Table\,\ref{t:tab1}) and shown in Figures\,\ref{f:fig3}--\ref{f:fig5}. We start with a case where the Lare3D rope emerges relatively far away from the TDm rope (Run 1), and then successively decrease the distance between the two ropes by changing the location of the emerging rope center along the $\theta$ direction (Runs 2--4). In Run 4, the emergence takes place exactly below the TDm rope. Note that Runs 1--3 correspond to cases termed ``unfavorable for reconnection'' by \cite{feynman95}, while Run 4 would be favorable in their terminology, as they considered all cases of emergence within a filament channel as favorable for reconnection, regardless of the magnetic orientation of the emerging flux (see Section\,\ref{s:intro}). 

%

%

%
\underline{Run 1:} In this run the flux emergence is centered 59\,Mm from the TDm center. As the new rope emerges, a current layer forms between its periphery and the coronal background field, see Figure\,\ref{f:fig3}(b). Due to the chosen orientation of the Lare3D rope, this layer forms at the far side of the emerging flux (as seen from the position of the TDm rope). Due to this relatively large distance of the current layer from the TDm rope, reconnection across this layer involves only very high-arching background field lines, which should not have any significant effect on the stability of the TDm rope. However, as can be seen by comparison with the initial configuration shown in Figure\,\ref{f:fig3}(a), the field directly overlying the TDm rope contracts considerably, increasing the magnetic tension above the rope. This is predominantly a consequence of the changing boundary conditions, as discussed in \cite{ding08} (see Section\,\ref{s:intro}). Because of the gradually increasing tension, the TDm rope is slowly pushed downward. The rope also leans slightly to the left, due to the repulsion of the TDm and Lare3D flux-rope currents, or, equivalently, the increase of magnetic pressure in the field that separates the two ropes. No eruption of the TDm rope occurs in this run. We note that the Lare3D rope erupts shortly after the time shown in the figure, but the eruption is directed away from the TDm rope and, therefore, does not significantly affect the latter. An eruption of the Lare3D rope occurs around the same time in almost all of our simulations, and we discuss it for those cases where it plays a significant role for the evolution of the configuration. 

%
\underline{Run 2:} Here the emergence is imposed closer to the TDm rope, 45\,Mm from the TDm center---this is the case for which the photospheric field is shown in Figure\,\ref{f:fig2}(a). The resulting evolution is very similar to Run 1, except that the expansion of the emerging flux is somewhat weaker (since the background field into which the Lare3D rope emerges is somewhat stronger), and the TDm rope leans a bit more to the side (since the repulsion of the flux-rope currents is stronger). Eventually, as can be seen in Figure\,\ref{f:fig3}(c), the TDm rope settles at a height that is slightly larger than in Run 1. 

%
\underline{Run 3:} In this run, we place the Lare3D rope even closer to the TDm rope, 31\,Mm from the TDm center. As in the previous two runs, the reconnection in the current layer involves only high-arching field lines of the background field, so its influence on the evolution is still negligible. As before, the intrusion of new flux at the boundary initially leads to a contraction of the field lines that are located right above the TDm rope. However, due to the proximity of the emerging Lare3D rope to the TDm rope, the repulsion of the flux-rope currents soon becomes the dominant effect, similar to the ``bounce'' cases in \cite{linton01}, and no significant downward movement of the TDm rope takes place. Instead, the rope is continuously pushed upwards and to the left by the emerging flux. Finally, at around the time shown in Figure\,\ref{f:fig3}(d), an eruption of the TDm rope commences, driven presumably by a combination of (flare-)reconnection below the rising TDm rope and the TI (see Section\,\ref{s:intro}), which is expected to occur once a flux rope reaches a critical height above the surface. The same driving mechanism(s) can be assumed to be at work in the TDm eruptions occurring in Runs 4 and 5 below.

%
\underline{Run 4:} In the final run of this series we center the emerging flux exactly below the TDm rope. Due to a strong direct interaction between the Lare3D and TDm fields, this leads to an evolution that is rather different from Runs 1--3. As can be seen in Figure\,\ref{f:fig3}(e), the TDm flux rope erupts in this run as well. Compared to Run 3 the eruption starts earlier ($t \approx 3$ vs. $t \approx 6$) and is more violent (the maximum kinetic energy in Run 4 is more than two times larger than in Run 3). 

Figure\,\ref{f:fig4} illustrates the evolution of the system in Run 4 in more detail. The top panel in Figure\,\ref{f:fig4}(a) shows the configuration at $t=1$, after the relaxation of the TDm configuration, before flux emergence is imposed (cf. Figure\,\ref{f:fig3}(a)). The TDm configuration we are using here has a so-called ``normal polarity'' configuration, i.e., a small arcade is present below the flux rope (see also the magnified view in the bottom panel of Figure\,\ref{f:fig4}(a)). The light-green field lines show the core of the TDm rope; its full cross-sectional extension is approximately outlined by the transparent electric currents in the figure. The orange field lines emanate from the main AR polarities close to the PIL, arch over one half of the TDm rope, and connect to the rope footprint of opposite polarity. As one moves towards the centers of the main polarities, the shear of these field lines (the component parallel to the PIL) continuously decreases, until the unsheared, light-blue field lines overlying the central part of the TDm rope are reached. All field lines below the core of the TDm rope are colored by magnetic energy density. For example, the long, largely horizontal field lines below the light-green ones are located in the bottom part of the TDm rope. 

As the new flux emerges, it first encounters the small arcade below the TDm rope. Due to the chosen orientation of the Lare3D rope, a quadrupolar flux distribution develops, i.e., the emerging field lines and the field lines in the small arcade are largely anti-parallel, and the arcade is, therefore, quickly ``reconnected away'' by the emerging flux. There is some downward movement of sections of the overlying field during this initial evolution, but this rather small contraction is reversed early on in the evolution ($t \gtrsim 2.5$).   

Figure\,\ref{f:fig4}(b) shows the system at $t=3.5$ (1 hour after the onset of emergence), when a significant amount of flux has emerged (cf. Figure\,\ref{f:fig1}(b)) and expanded in the corona. Note that, once the small arcade is reconnected away, the angle between the emerging and the pre-existing field lines is much smaller at the top of the emerging flux system than at its flanks. Consequently, two current layers form at these flanks \citep[highlighted by black lines in the bottom panel; see also Figure\,3(c) in][]{ding08}, across which reconnection takes place between the outermost emerging field lines and the sheared orange field lines. A pair of such reconnection events has occurred in Figure\,\ref{f:fig4}(b) on either side of the emerging flux, creating the undulating orange field line shown there. The overlying light-blue field lines do not get involved in this reconnection. No strong current layer forms at the top of the emerging flux and, therefore, no significant reconnection takes place between the emerging flux and the bottom part of the TDm rope either. The newly reconnected orange field lines expand upwards as they try to straighten out, thereby aiding the slow rise of the TDm rope, which is described next.

Due to the absence of reconnection between the emerging flux and the TDm rope, the magnetic pressure below the rope gradually increases, as illustrated in Figure\,\ref{f:fig4}(c). The top panel shows $B_\theta$, the field component perpendicular to the TDm and Lare3D flux-rope axes. Prior to emergence ($t=1$), the two locations where $B_\theta$ changes sign outline the positions of the X-line below the TDm rope and axis of the rope, respectively. As the new flux emerges and expands in the corona ($t=2.5$), the magnetic pressure (central panel) below the TDm rope strongly increases and the bottom part of the rope is pushed upward while its axis hardly moves (see vertical lines). This pile-up of magnetic pressure below the rope (corresponding to the repulsion of the two current systems) finally starts to push the whole TDm rope upwards until the rope becomes unstable and erupts (bottom panel; $t=3.5$). The reconnection at the sides of the emerging flux, discussed above, may aid the rise of the flux rope to some degree, but the main effect driving the rope to eruption seems to come from the repulsion of the currents. Again, this is similar in nature to the ``bounce'' effect described in \cite{linton01}. Since the two flux-rope currents are closer to one another than in Run 3, the eruption of the TDm rope sets in much earlier (see Figure\,\ref{f:fig3}). 

\begin{figure*}[t]
\centering
\includegraphics[width=0.95\linewidth]{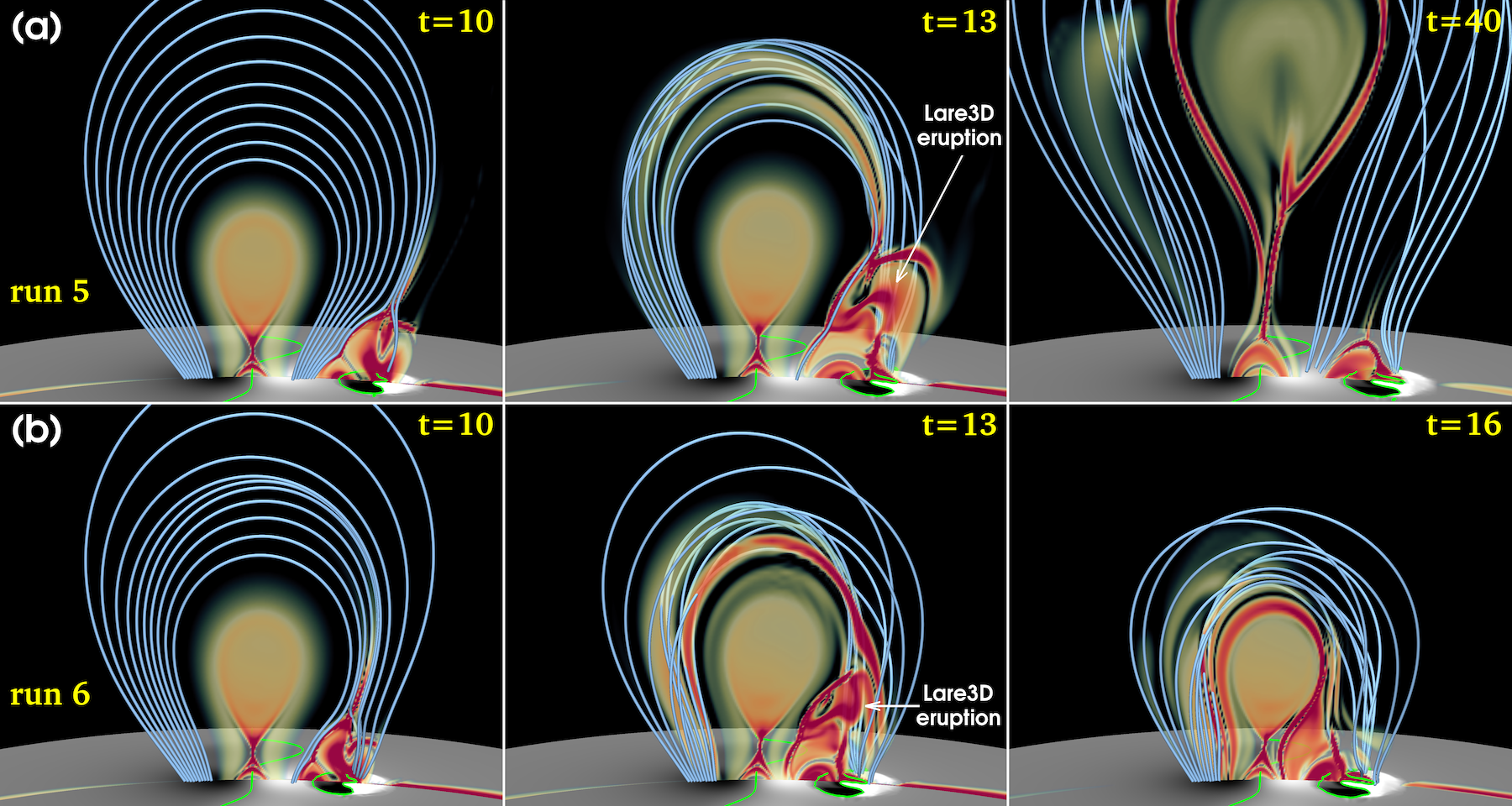}
\caption{
Same as Figure\,\ref{f:fig3}, for Run 5 (a) and Run 6 (b), at three consecutive times. The initial state is the same as in Figure\,\ref{f:fig3}(a), but here the orientation of the emerging flux is reversed with respect to the runs shown in Figure\,\ref{f:fig3} (see text for details).  
}
\label{f:fig6}
\end{figure*}

Figure\,\ref{f:fig5} highlights several interesting features of the subsequent evolution of this simulation. Panel (a) shows the eruption of the TDm rope. It can be seen that an elongated vertical (flare) current layer forms below the erupting rope, as expected from the ``standard model'' of solar eruptions \citep[e.g.,][]{janvier14a}. Panel (b) then shows the subsequent, ``homologous'' eruption of the Lare3D rope. Panel (c) shows that, over the course of the emergence process, two so-called ``conjoined flux ropes'' (CFRs) are created at the periphery of the emerging flux-region. A CFR consists of two flux ropes with opposite signs of twist that are ``connected'' via a spiral null point \citep[e.g.,][]{lau90,parnell96,wyper14,olshevsky16,titov17}. The CFRs likely form here as a result of the tearing instability occurring in the current layers that have formed between the emerging and pre-existing flux (see Figure\,6 and last paragraph of Section\,2.1.2 in \citealt{patsourakos20} and references therein). On the Sun, CFRs may provide one possible magnetic configuration for harboring filaments at the periphery of ARs. A detailed description of their formation process in flux-emergence simulations will be provided in a forthcoming publication (T\"or\"ok {\em et al.}, in preparation). In summary, our results indicate that reconnection between the emerging flux and the ambient (overlying) TDm field has a negligible effect on the stability of the TDm rope in all four simulations of this series. The evolution is rather driven by the competing effects of the contraction of the ambient field due to the intrusion of new flux (which moves the TDm rope downward) and the repulsion of the flux-rope currents (which moves the TDm rope upward). The respective contributions of these effects depend on the distance between the TDm rope and the emerging flux: the smaller the distance, the more dominant is the repulsion. For sufficiently small distances, this repulsion can lift the TDm rope to an unstable height and trigger its eruption. The eruptive versus non-eruptive behavior of Runs 1--4 are in line with the findings by \cite{feynman95}. Runs 1--3 correspond to their ``unfavorable'' configurations, for which those authors report that two out of five cases erupted. Run 4 is a ``favorable'' configuration in their terminology, for which they found all 17 observed cases erupted.

%
\subsection{Co-aligned axial current \& counter-aligned axial field}
\label{ss:co-j_counter-B}

In our second series (Runs 5--8), we consider configurations in which the axial currents of the TDm and Lare3D ropes are co-aligned, while their axial fields are counter-aligned. This is achieved by simply reversing the orientation of the Lare3D rope (Figure\,\ref{f:fig2}(b)). We center the flux emergence at the same locations as in Runs 1--4 (see Table\,1). All runs of this series correspond to what \cite{feynman95} would have termed ``favorable for reconnection''.  

%
\underline{Run 5:} This run is shown in Figure\,\ref{f:fig6}(a). As in Run 1, the emerging flux is centered 59 Mm from the TDm center. Due to the opposite orientation of the emerging flux, a quadrupolar distribution of the surface flux develops, and the current layer now forms {\em between} the emerging AR and the TDm rope. Reconnection across this layer affects overlying field lines at significantly lower heights than in Run 1. This reconnection transfers foot points of overlying field lines to the positive (white) polarity of the emerging flux, thereby increasing their length and reducing the magnetic tension. Moreover, the changes of the magnetic surface distribution now lead to an expansion of the overlying field. Consequently, at $t \approx 3$, the latter starts to expand notably, and the TDm rope begins to rise slowly. 

The left panel in Figure\,\ref{f:fig6}(a) shows the configuration at $t=10$. It can be seen that the emerging flux has expanded significantly, and an arcade of reconnected field lines has formed below the current layer, harboring a CFR. The TDm rope slightly leans towards the right, which can be attributed to the slightly asymmetric expansion of the overlying field, and possibly also to the attraction of the (oppositely directed) flux-rope currents, even though, at this relatively large distance between the ropes, the attraction should be rather weak.

\begin{figure*}[t]
\centering
\includegraphics[width=1.\linewidth]{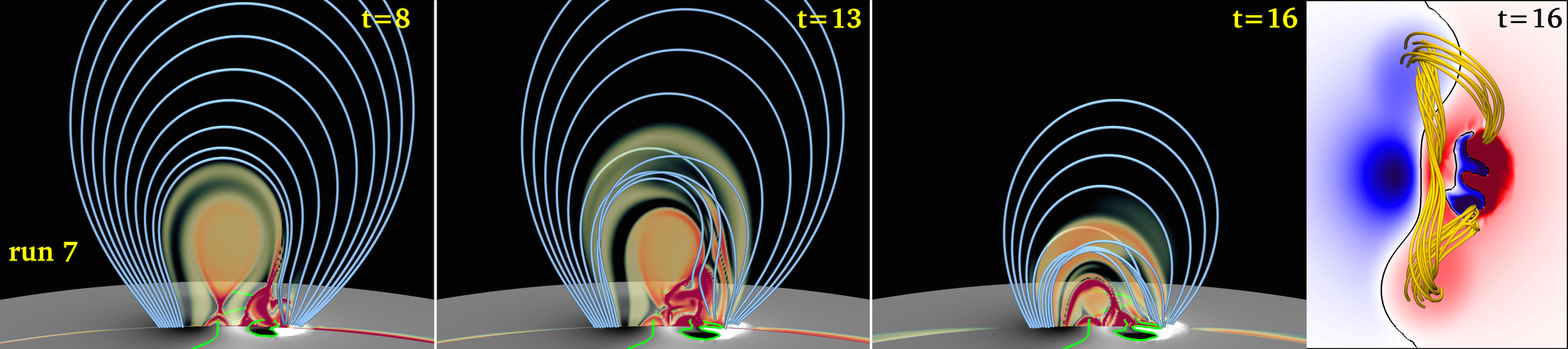}
\caption{
(a--c) Same as Figure\,\ref{f:fig3}, for Run 7, at three consecutive times. The initial state is the same as in Figure\,\ref{f:fig3}(a), but here the orientation of the emerging flux is reversed with respect to the runs shown in Figure\,\ref{f:fig3}. 
(d) Top-down view showing magnetic connections after the interaction of the flux ropes (see text for details). $B_z(r=1)$ is shown here in red/blue colors for better contrast. 
}
\label{f:fig7}
\end{figure*}

As in the previous runs, the Lare3D rope erupts during its emergence, starting at $t \approx 10$. In contrast to Runs 1--3, where the eruption was always directed away from the TDm rope, the Lare3D rope erupts here in a largely radial direction. Therefore, it strongly perturbs and reconnects with a part of the ambient flux close to the TDm rope, though the TDm rope itself does not get involved in this reconnection (central panel in Figure\,\ref{f:fig6}(a)). The TDm rope now leans more to the side, due to the increased attraction of the two approaching flux-rope currents. The interaction of the Lare3D rope with the ambient field leads to a contraction of the latter, which temporary slows down the rise of the TDm rope. However, shortly after this perturbation, its rise accelerates again and the TDm rope finally erupts (right panel in Figure\,\ref{f:fig6}(a)). Hence, the Lare3D eruption disturbs and delays the rise of the TDm rope, but does not suppress its eruption. The eventual eruption of the TDm rope evolves slower than in Run 3, which is most likely because the rope has to overcome the contraction of the overlying field that is not present in Run 3.

%
\underline{Run 6:} Next we center the emerging flux closer to the TDm rope, 45 Mm from the TDm center (as in Run 2). This run is shown in Figure\,\ref{f:fig6}(b). The initial evolution is very similar to Run 5; the TDm rope starts to rise slowly at $t \approx 3$.  Due to its closer vicinity to the TDm AR, the emerging flux expands less in the corona and reconnects with somewhat lower-arching ambient field lines, and the TDm rope leans slightly more to the right ($t=10$). Since the Lare3D rope now erupts closer to the TDm rope than in Run 5 ($t=13$), the resulting reconnection leads to a contraction of field lines that arch directly above the TDm rope, and the attraction of the flux-rope currents is stronger. Both effects pull the TDm rope downwards and manage to overcome the upward motion of the TDm rope. Eventually, the system settles into the state similar to the one shown at $t=16$, without producing a TDm eruption. 

We note, however, that the TDm rope started off rising and behaving as if it would erupt until the eruption of the Lare3D rope disturbed its rise. This led us to hypothesize that the TDm rope might erupt if less of the Lare3D rope were allowed to emerge. We tested this hypothesis in a separate run, where we avoided the eruption of the Lare3D rope by stopping the flux emergence earlier (at $t=9$). This, however, still failed to generate a TDm eruption. While stopping the emergence suppressed the eruption of the Lare3D rope, it also meant that the tension of the field lines overlying the TDm rope was not reduced by as much, and so the TDm rope also did not erupt. However, we postulate that, if the emergence had continued in our original run without the Lare3D rope erupting, an eruption of the TDm rope would have commenced after a relatively short time, as the latter showed clear indications of an impeding eruption before the interaction with the Lare3D rope suppressed it. Demonstrating this would require a different setup, which we leave for a future study.

%
\underline{Run 7:}
In this run, we place the Lare3D rope center 31\,Mm from the TDm center (as in Run 3). Just as in Runs 5 and 6, we initially see an expansion of the overlying field, as expected. However, in contrast to those two runs, the expansion does not trigger a pronounced slow rise of the TDm rope here. Rather, due to the closer proximity of the emerging Lare3D rope and the resulting stronger attraction of the flux-rope currents, the TDm rope leans to the side early on and more strongly, so that the two ropes approach each other more closely than in Runs 5 and 6 (Figure\,\ref{f:fig7}; $t=8$). 

As in all previous runs, the Lare3D rope erupts after some time. In this case, in contrast to the previous runs of this series, this yields (direct) reconnection between the two flux ropes, after the small amount of separating ambient flux is quickly removed (Figure\,\ref{f:fig7}; $t=13$). This reconnection is of the same nature as the slingshot reconnection described in \cite{linton01} and \cite{torok11}, and it leads to an exchange of the footprint connectivities of the two ropes, which is accompanied by a strong shrinking of the ambient field during this highly dynamic evolution (Figure\,\ref{f:fig7}; $t=16$). The exchange of the footprint connectivities is only partial, as the reconnection did not fully split the TDm rope by the end of the simulation ($t=16$). Note that the direct interaction of the two ropes is forced here predominantly by the eruption of the Lare3D rope. We postulate that if the emergence had continued without an eruption, the connectivity changes would likely still have occurred (since the rope currents attract each other), albeit at a much slower rate.

\begin{figure}[t]
\centering
\includegraphics[width=0.96\linewidth]{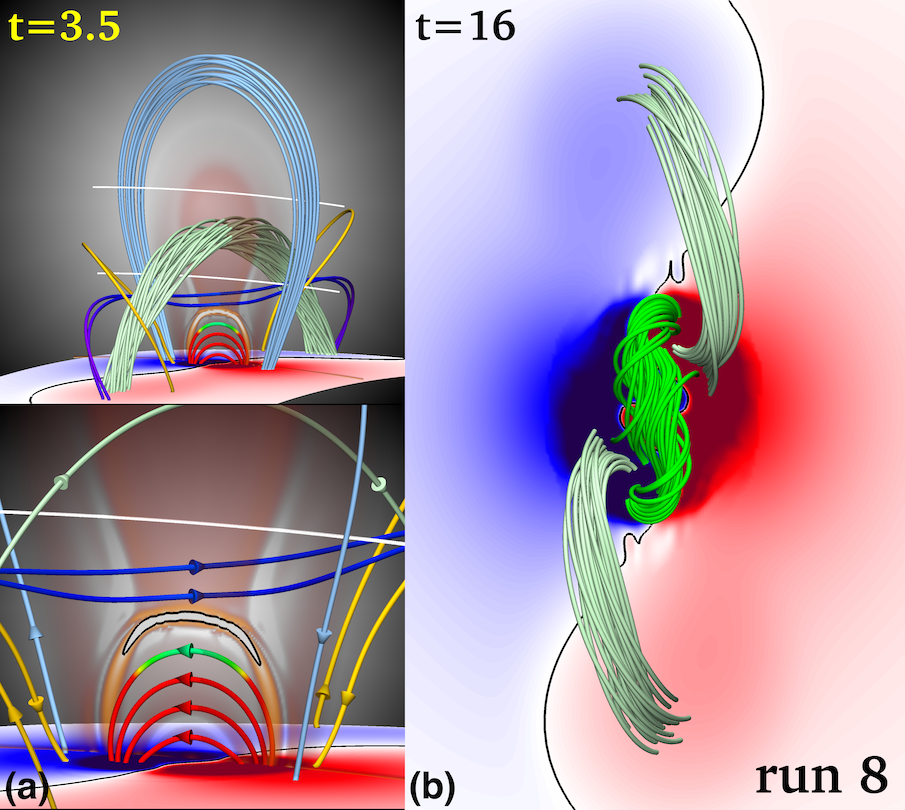}
\caption{
(a) Same as Figure\,\ref{f:fig4}(b) for Run 8. Here the emergence leads to a bipolar flux distribution in the AR center, and a current layer forms on top of the emerging flux rather than at its flanks.
(b) Top view on the AR at the end of the simulation. Reconnection across the current layer has led to a full splitting of the TDm rope. The dark green field lines outline the core of the emerged flux.
} 
\label{f:fig8}
\end{figure}

%
\underline{Run 8:} Finally, we place the emerging flux region directly below the TDm rope, as in Run 4. In contrast to that run, for which the emerging flux had the opposite orientation, the emergence here leads to a bipolar configuration underneath the TDm rope (Figure\,\ref{f:fig8}). As a result, the outer emerging field lines now make a relatively small angle with the surrounding sheared field lines, represented by the two orange field lines in Figure\,\ref{f:fig8}(a), which start at the same locations at the bottom boundary as in Figure\,\ref{f:fig4}. Instead, these outer emerging field lines make a large angle (of more than 90 degrees) with the field lines at the bottom of the TDm rope. Consequently, as highlighted by the black contours in the bottom panel of Figure\,\ref{f:fig8}(a), a strong current layer now forms primarily at the top of the emerging flux rather than at its flanks (cf. Figure\,\ref{f:fig4}(b)). Specifically, the reconnection between the orange field lines and the emerging flux, which aided the rise of the TDm rope in Run 4, is now largely, if not completely, absent. Therefore, as the emerging flux expands into the TDm configuration, it reconnects directly and efficiently with the TDm rope. We did not notice any slow rise of the TDm rope prior to the onset of this reconnection. Eventually, since the TDm rope does not move upward and contains less flux than the emerging Lare3D rope, a complete splitting of the TDm rope occurs, as depicted in Figure\,\ref{f:fig8}(b).

Evidence for partial or full flux-rope ``splitting'', as seen in Runs 7 and 8, can occasionally be observed on the Sun if filament material is present. For example, \cite{li.t15} and \cite{dacie18} report observations of apparent filament splitting during the emergence of a compact flux region underneath or in the vicinity of a pre-existing filament, respectively.

In summary, the configurations obtained in our second simulation series differ from the first series (Runs 1--4) in several respects. Firstly, in Runs 5--7, where emergence occurs next to the TDm rope, the current layer forms {\em in between} the interacting flux systems, so that any changes in the overlying field lines due to reconnection occur at lower heights than in Runs 1--3 and, therefore, have a more pronounced effect. Secondly, the intrusion of new flux now leads to an {\em expansion} of the ambient flux, helping the TDm rope to rise. Finally, the flux-rope currents now {\em attract} each other, counteracting such a rise. The relative contributions of these three effects depend, again, on the distance between the emerging flux and the TDm rope. For the two cases where the emergence is further away from the TDm rope (Runs 5 and 6), the former two effects seem to dominate, and the TDm rope moves upward and erupts (in Run 6, however, the eruption is likely suppressed by the erupting Lare3D rope). For the case where the emergence is close by (Run 7), the current-attraction dominates and the TDm rope is slowly pulled towards the emerging flux. When the Lare3D rope erupts into the TDm rope, strong reconnection between the flux systems ensues and the TDm rope is partly destroyed (i.e., it partially splits). In Run 8, where emergence occurs below the TDm rope, the strongest currents form at the {\em top} of the emerging flux, rather than at its flanks. This leads to direct and strong reconnection between the TDm rope and the emerging flux, culmination in the full destruction (splitting) of the TDm rope. In the terminology of \cite{feynman95}, all four configurations in this series are ``favorable'' for eruption, yet we find an eruption (or indication of eruption) only in two cases. The other two cases, where current-attraction hinders an eruption from occurring in our configurations, are contrary to their findings. 

%
%
\section{Discussion}
\label{s:dis}

We have presented a series of $\beta=0$ MHD simulations of localized, boundary-driven (``kinematic'') emergence of magnetic flux in the vicinity of (or below) a pre-existing filament channel. Our work extends the previous analytic and numerical studies mentioned in Section\,\ref{s:intro} by considering, for the first time, a pre-existing 3D flux rope in the corona. The flux emergence was driven by imposing a time series of slices, extracted from a dynamic simulation of the emergence of a sub-photospheric flux rope into the corona, at the bottom boundary of our simulation domain. We considered two main sets of cases. In the first set the axial currents of the coronal and emerging flux ropes are counter-aligned, while the axial magnetic fields are co-aligned. In the second case, the axial currents are co-aligned while the axial magnetic fields are counter-aligned. For each set, we varied the location of the emerging flux region relative to the location of the coronal flux rope. The full set constitutes five cases in which the orientation of the emerging flux was ``favorable for reconnection'' as proposed in the pioneering work by \cite{feynman95}, and three cases were the orientation was ``unfavorable''.  Our results confirm the overall main conclusion obtained in these studies: whether or not an eruption of the pre-existing flux occurs in such a system depends strongly on parameters such as the location and magnetic orientation of the emerging flux.

We identified three different mechanisms that can lead to eruptions, namely (i) reconnection displacing foot points of field lines overlying the pre-existing flux rope, (ii) expansion of the ambient field due to the intrusion of new flux at the boundary, and (iii) repulsion of the (axial) electric currents in the pre-existing and newly emerging flux systems. For the cases we have studied here, their respective contributions (if any) depend on both the distance between and relative magnetic orientation of the pre-existing and newly emerging flux. Overall, current-repulsion in the case of emergence below the pre-existing rope (Run 4) seems to be the most efficient mechanism, as in that case the eruption takes place much earlier and is more impulsive than in all other eruptive cases. However, this effect may depend significantly on the strength of the emerging flux. Consequently, another important parameter, not studied here, should be the relative strength (flux content) of the two interacting flux systems. Furthermore, the question of how far away the pre-existing flux rope is from the onset condition of the TI (i.e., the critical height at which the instability is triggered) is sure to be relevant.       

Six of our eight runs are in agreement with the findings by \cite{feynman95}, who observed filament eruptions in two out of five cases with ``unfavorable'' orientation of the emerging flux, and in all 17 cases with ``favorable'' orientation. However, two of our runs that have a favorable orientation for eruption in the \cite{feynman95} classification did not lead to an eruption. In the latter two cases, the emergence occurred close to the coronal rope, so that current-attraction and subsequent slingshot reconnection (aided in one case by the eruption of the emerging flux), led to the destruction (``splitting'') of the coronal rope. There may be several reasons for this disagreement with \cite{feynman95}. First, while all favorable cases reported by those authors exhibited the same behavior, their sample of 17 events is relatively small. The case reported by \cite{li.t15}, while apparently being rare, suggests that emergence within a filament channel does not necessarily lead to an eruption. This conclusion is also supported by the simulations of \cite{chen00}, \cite{xu05}, and \cite{kusano12}. Further studies based on larger samples, utilizing the present observational capabilities of, e.g., the {\em Solar Dynamics Observatory} \citep[{\em SDO};][]{pesnell12}, are needed to obtain improved estimates of that fraction. 

Second, the electric current in our emerging flux is relatively strong compared to the current in the TDm rope. A weaker emerging current will likely not lead to such a strong attraction of the currents (and subsequent reconnection) as in the two runs in question, potentially allowing an eruption of the coronal flux rope to happen. 

Third, we have only considered two orientations of the emerging flux in our runs namely an emerging flux rope with its axial magnetic field parallel to or antiparallel to that of the coronal rope. It may well be that, despite potential attraction of the currents, a large fraction of intermediate relative orientations of the two axial magnetic fields may lead to eruptions. While, at first glance, this is not supported by the results of \cite{kusano12}, where the majority of relative orientations between emerging and pre-existing flux did {\em not} lead to a large-scale eruption (even though a stabilizing, poloidal (``strapping'') field was absent in their setup), it has to be kept in mind that in their simulations a large-scale erupting flux rope first had to be {\em formed} by reconnection. For a fully developed pre-existing flux rope, as in our case, the situation may be very different. 

Due to the complexity of the system, we could only consider a limited set of possible interactions here. Future work will investigate potential changes to the evolution with the eruption of the emerging flux suppressed, the interaction of flux ropes with the same sign of twist (handedness), as well as system parameters that were not considered here, such as the relative flux contents (strengths) of the emerging and pre-existing flux systems, the speed of the emergence, the intermediate orientations of the initial Lare3D flux-rope axis relative to the TDm flux-rope axis, and the initial distance of the TDm rope axis apex from the critical height for the onset of the TI. In the TDm configuration used here, the flux-rope axis adopts an almost semi-circular geometry after the initial relaxation (see Figure\,\ref{f:fig4}(a)), implying that the rope is relatively close to TI onset. Flatter flux ropes (for the same AR background field) can readily be produced by reducing the initial axial current \citep[see, e.g., Figure\,4 in][]{titov14}.

In the future, we also plan to include the continuity and energy equations into our calculations. As discussed in \S\,\ref{ss:MAS}, the $\beta=0$ approximation used here is well justified for the low corona, and we tried to account for pressure stratification at and near the photosphere by employing driving conditions extracted from a fully dynamical flux-emergence simulation (see \S\,\ref{ss:bc}). While we do not expect a fundamentally different behavior of the system, the quantitative properties of the evolution (such as the onset times and rates of reconnection) will likely differ and may, to some extent, change the conditions under which the TDm rope erupts.

\acknowledgments
We thank the referee for  constructive and helpful comments, and Jon A. Linker for helpful discussions on the numerics. This work was supported by Naval Research Lab (NRL) contract N0017319C2003 to Predictive Science Inc., a subcontract of NASA LWS grant 80HQTR19T0084 to NRL. C.D., R.L., T.T., and V.S.T. acknowledge support by NASA's HTMS program (award No.\ 80NSSC20K1274). C.D., T.T., and V.S.T. were additionally funded by NSF's PREEVENTS program (award No.\ ICER-1854790) and by NASA's HSR program (award No.\ 80NSSC20K1317). T.T. and V.S.T. further acknowledge funding by NASA's HGI program (award No.\ 80NSSC19K0263). M.G.L. was in part supported by the Office of Naval Research. J.E.L was supported by NASA ISFM Work Packages and by NASA's NNH21ZDA001N-HSR program. Computational resources were provided by the NSF supported Texas Advanced Computing Center (TACC) in Austin and the NASA Advanced Supercomputing Division (NAS) at Ames Research Center. The Lare3D simulations were performed with a grant of computing time from the DoD HPC program.

\newpage

\bibliographystyle{yahapj}
\bibliography{torok}
  
\listofchanges
\end{document}